\documentclass[a4paper,10pt]{article}
\usepackage{graphicx}

\addtocounter{page}{-1}
\begin{document}
\pagestyle{empty}
\parindent 8mm
\pagestyle{empty}

\vspace*{2cm}
\title{The Phase Structure in the Evolution of Globally coupled Spatial Prisoner's Dilemma}
\author{Norihito Toyota and Shota Hayakawa}
\date{Hokkaido Information University,
59-2 Nishinopporo Ebetsu City \\
E-mail:toyota@do-johodai.ac.jp
}

\maketitle
\baselineskip 5mm
\pagestyle{empty}
\begin{abstract}
We propose two new evolutionary rules that is not mimic evolution of strategies based on the spatial Prisoner's Dilemma (PD). 
The former follows  the selfish evolutionary rule and then the coexistence phase appears with weak phase transition with mutations. 
The second rule is that of globally coupled spatial PD where the strategy of an agent is taken  so that the  total payoff of the whole system increases.  
Then it is natulally suspected that all agents will become cooperators in general.  We, however, find that situations is rather complerx. 
We find two critical points within the usual conditions in the parameters included the payoff matrix in the PD. 
Further as we break the condition without losing PD condition, we meet two more  essential curitical points. 
We discuss the areas between these critical points in the parameter space and 
uncover the various properties by simulating on computer. 
We show that in some parameter regions, strange properties, that is, like chaotic behaveor appear and the coexistence between defectors and cooperators appears in the large parameter region. Thus we  point out that usual non-essential condition among the parameters in the payoff matrix imposed on the PD privent from recovering rich structures in the evolutionally spatial PS.
\end{abstract}
\bf{key words:} 
\it{Prisoner's Dilemma, Evolutionary Game, Spational Game, Phase Structure, Globally Coupled, Critical Point}

\rm
\normalsize
\baselineskip 6mm
\pagestyle{empty}
\thispagestyle{empty}

\renewcommand{\thefootnote}{\fnsymbol{footnote}} 
\rm
\normalsize
\baselineskip 6mm

\section{Introduction}
Traditionally differential equations have been applied to understand various phenomena in the nature and the artificial, especially physical,  phenomena, since they were used to construct classical dynamics by I. Newton. 
It, however,  seems that some phenomena are rather relevant to be studied as they follow some descrete dynamics.
 They may include biological, social, ecological phenomena and so on, which have a close relationship to "information".
 The game theory among many aproaches give one of interesting view of them. 
In this article we discuss the Prisoner's Dilemma game (PD), which is the most famous game  and has substantial content among many sorts of games. 

PD is symmetric game played by two agents and the payoff function acquired in each strategy is discribed by matrix form. 
One shot game of PD can be analyzed analytically \cite{Okad}.   
Many researches have been made to resolve the Dilemma, $pareto \; optimal \neq Nash\; equilibrium$, and how can we obtain cooperation rationally.  
Based on PD, iterated games have been studied and Axelrod \cite{Axel} has shown that cooperation can emerge as a norm in a society comprised of individuals with selfish motives. 
Moreover spatial version of PG has been discussed so that a defection is to be only evolutionary stable strategy (ESS) \cite{Smit} if each agent interacts with any other agent. This aspect drastically changes if a spatial structure of the population is considered. If the interaction between agents is locally restricted to their neighbors \cite{Nowa}, a stable coexistence between cooperators and defectors become possible under certain conditions \cite{Nowa}\cite{Nowak}. 
In the case, one agent plays PD with their neighbors and in next step the agent take the same strategy of the agent that acquired hightest payoff among the neighbors,  which reflects to Dawin's theory, sometimes "including mutation"\cite{Fort}. 
Then it is assumed that all agents play at same time and follow the same way. 
Recently the evolution of the spatio-structured PD has been systematically explored in details in Ref.\cite{Schw}.

We here discuss two other evolutionary rules within the framework of the spatio-structured PD where agents interact locally, in which the evolutionary rule entices  the cooperator and defector to coexistence phase. 
One way to realize it is to change his(her) strategy when the agent would get more payoff if he(she) had changed the strategy. 
This  selfish rule, however, leads to a trivial result, since PD has a dominant strategy.  
Second one is to change the strategy like totalitarian, that is, if the total payoff yielded in whole system increases when an agent changes the strategy. 
We call this evolutionary rule the grobally coupled evolution. 
This evolutionary role is expected to lead to full cooperative strategy but it do not so. We found 4 critical points analytically, and perform computer simulations in the parameter regions bounded on the critical points.

 After Introduction, section 1, we discuss our evolutionary strategy and the critical points  analytically in the section 2. In the section 3 simulation results are obtained based on the results in the section 2. The final section 4 is devoted to summary. 

\section{Evolutionary Rule and Analytical Study}
Spationally structured PD is first explained briefly. 
Then we discuss two evolutionary rules that are not considered thus far and study them analytically. 

\subsection{Spational Structured PD as Cellular Automaton}
Two types of agents, cooperators (C) and defectors (D) are considered.
The distinction is made by means of a variable $\theta \in \{-1,1\}$, where $1$ refers to cooperation and $-1$ refers to defection. 
An individual agent $i$ is indicated by $\theta _i$. 
$N$ is total number of agents and the fraction of cooperators and defectors are given by 
\begin{eqnarray}
N&=& \sum_\theta N_\theta =N_{-1}+ N_1=const,\\
f_{\theta}&=&\frac{N_\theta}{N};\;\;\; f_{-1}+f_{1}=1.
\end{eqnarray}
The spatial distribution of the agents are considered as a two dimensional cellular automaton (CA) consisting of $N=n^2$ cells, where each cell is identified by the index $i \in N$ refering its spacial position and the state $\theta _i$ sits on the cell. The state space of all possible configurations is of the order $2^N$. 

We asumme that each agent $i$ has two options, C ($\theta =1$) or D ($\theta =-1$). 
Playing with 4 neighbors of $i$, the action chosen by $i$ and actions chosen by the 4 agents only affect the total payoff (Hamiltonian). 
We assume that an agent $i$ simultaneously plays  with 4 agents with a same strategy,  and so the game essentially reduces to 2-person game. Then the two person game can be  described by a payoff matrix and it is obtained by the following 
Table 1 for PD.

\begin{table}[h]\centering
\caption{Payoff table for PD. The left and right variables in the parenthesis show the payoffs of the agent $i$ and $j$, respectively. }
\begin{tabular}{|c|c|c|} \hline
  &\makebox[15mm]{C$_j$} & \makebox[15mm]{D$_{j}$} \\ \hline
C$_{i}$ & ($R,R$) & ($S,T$) \\ \hline
D$_{i}$ & ($T,S$) & ($P,P$) \\ \hline
\end{tabular}
\end{table}
In PD, the payoffs have to fulfill the following condition:
\begin{equation}
T>R>P>S,
\end{equation}
and usually for iteration of game an additional conndition is imposed:
\begin{equation}
2 R> S+T.
\end{equation}
The second condition, however, is considered to have not any essential meaning for the structured spatial game. 
Later the breakdown of this condition (4) induces some interesting phenomena in spatio PD. 
A simple analysis show that strategy D is so-colled ESS in a one-shot PD \cite{Smit}. 
It is also known that the pareto optimal strategy is for both agents to take C but Nash solution is  for both agents to take D, which leads to famous "Dilemma".       
   
In this paper we are interseted in the spacial effects of PD and assume that each agent interects only with the 4 agents of its neighborhood. In evolutionary game theory this is called a 4-person game.  
We here investigate the modified evolutionary rules, where players do not mimic the best strategy in its neighbors. 
We consider two extreme ways,
one obeys a selfish evolutionary rule and another is like totalitarian, where explict values of $T, S, R, P$ play important role apart from the usual PD.  

\subsection{Selfish Evolution}
Selfish evolution is to change strategy taken by an agent if the (target) agent get more payoff when he(she) changes the strategy, independent of other agents. Suppose that all the agents follow this rule and each agent updates its strategy in regular order. 
We see this case leads to a trivial result. 

In the simulation experiment of sec. 3 we include mutant agents with some probability with respect to Bolzmann distribution among regular agents that obey the selfish rule. This type agents will lead to the coexistence phase between C and D agents.          

\subsection{Globally Coupled Evolution}
As the second evolutionary model, we consider that if the total payoff, hamiltonian, yielded in whole system increases when an agent changes the strategy, the agent assumes to change its strategy. 
So an agents decide its strategy according to the total payoff of the whole system. Due to this property, we call this model "globally coupled". 
Notice that the evolution of the system depends on from which agent the strategy is renewed at each round, since by changing a strategy of each agent the Hamiltonian generally is updated and the next target agent decides its strategy based on the updated Hamiltonian. 
It means that the order of agent targeted is important and the evolutionary process is only determinstic under a order. 
We mainly discuss the case of the following ordering;
\[
\mbox[Do[ Do[ agent(i,j),\{i,1,n\}], \{j,1,n\}]]
\]
in Mathematica notation,  where $(i,j)$ shows the coordinate (row element, colum element) in latice like CA. 
A random ordering will be also considered latterly .
     
\subsection{Analytic study and critical points}
We analytically investigate globally coupled evolution of PD in this subsection. Let consider  the profit $P_i(C)$ that a cooperator agent $i$ encircled by $c$ cooperators and $d$ defectors; 
\begin{equation}
P_i(C)=cR+ dS, \;\;with\;\; c+d=4.
\end{equation}
If the agent change the strategy from C to D, the updated profit $P_i(D)$ of the agent is given by
 \begin{equation}
P_i(D)=cT+ dP,
\end{equation}
and the difference is 
\begin{equation}
\delta P(C\rightarrow D)=c(T-R) + d(P-S)>0,
\end{equation}
because of the equation (3).
Noticing that the increment $\delta h(\theta _j)$ of the profit of a neiborhood $j$ of the agent $i$ brought by this change is 
\begin{equation}
\delta h(\theta _j)=\left\{
\begin{array}{ll}
S-R& \mbox{ for } \theta_j=1\\
P-T& \mbox{ for } \theta_j=-1,
\end{array}
\right.
\end{equation}
the total $\delta h(\theta _j)$ resulted by 4 neighbors ($c$ cooperators and $d$ defectors)
is given by 
\begin{equation}
\Delta h= c(S-R)+d((P-T))<0.
\end{equation}
This reverse of the strategy of agent $i$ reflects the total profit (Hamiltonian) of the whole system and thus the resultant increment of the total Hamiltonian $\Delta H_{iC}[c,d]$ is given by
\begin{equation}
\Delta H_{iC}[c,d]= c(S+T-2R) + d(2P-S-T),
\end{equation}
where the first term in the right hand side is negative if the equation (4) is satisfied.  In contrast with this when  the agent $i$ change the strategy from D to C, the corresponding amount becomes 
\begin{equation}
\Delta H_{iD}[c,d]=  -\Delta H_{iC}[c,d]\equiv -\Delta H[c,d]
\end{equation}
by following the similar calculation. Explicitly they are written by
\begin{eqnarray}
\Delta H[0,4]&=& 4(2P-S-T),\\
\Delta H[1,3]&=& 6P -2(R+S+T),\\
\Delta H[2,2]&=& 4(P-R)<0 \;\; \mbox{from (3)},\\
\Delta H[3,1]&=& 2(P+ S+T- 3R)<0 \;\mbox{from (4)},\\
\Delta H[4,0]&=& 4(S+T-2R)<0 \;\; \mbox{from  (4)},
\end{eqnarray}
where (14) is negative by definition of PD, and (15) and (16) are negative because of the additional condition (4). Whether (12) and (13) are positive or not depends on the explicit parameters even when satisfying (3) and (4). 
This point is differnt from usual analysis of PD where explict values of $P,R,S,T$ are not essential but the order of the magunitudes plays only important role.  A standard set of parameters adopted by Okada \cite{Okad} is $\{P,R,S,T\}= \{-3, 5, -4,6\}$ and then all increments of the Hamiltonianturn are negative. 

However, whenever the condition $2P<S+T$ is satisfied, $\Delta H[0,4]$ and $\Delta H[1,3]$ come to be positive, and moreover (15) and (16) become possible to be positive when the the aditional condotion (4) is broken.  
To explain this situation, without losing the generality  we can choose $S$ and $T$ as variable parameters while $P$ and $ R$ are fixed to be the standard values.  
Namely as $S$  decreases much, $\Delta H[0,4]$ and $\Delta H[1,3]$  increase to be positive and as $T$  increases much, $\Delta H[3,1]$ and $\Delta H[4,0]$  increase to be positive. 
We find critical points in $S$ when $\Delta H[0,4]$ and(or) $\Delta H[1,3]$ become $0$, respectively. 
They turn out to be $S=-12$ and $S=-20$ whose values depend on the explicit values of $P$ and $R$ but this is not essential,  because the fact that there are two critical points in any values of them and the signs of $\Delta H[0,4]$ and $\Delta H[1,3]$ change preserves to be true at these points.  
There are equally two critical points in $T$, when $H[3,1]$ and $\Delta H[4,0]$ cross zero point. 
They are $T=14$ and $T=22$. Thus the parameter space divides into five regions and they summerized in the Fig.1. 
\begin{figure}[t]
\begin{center}

\includegraphics[width=10cm]{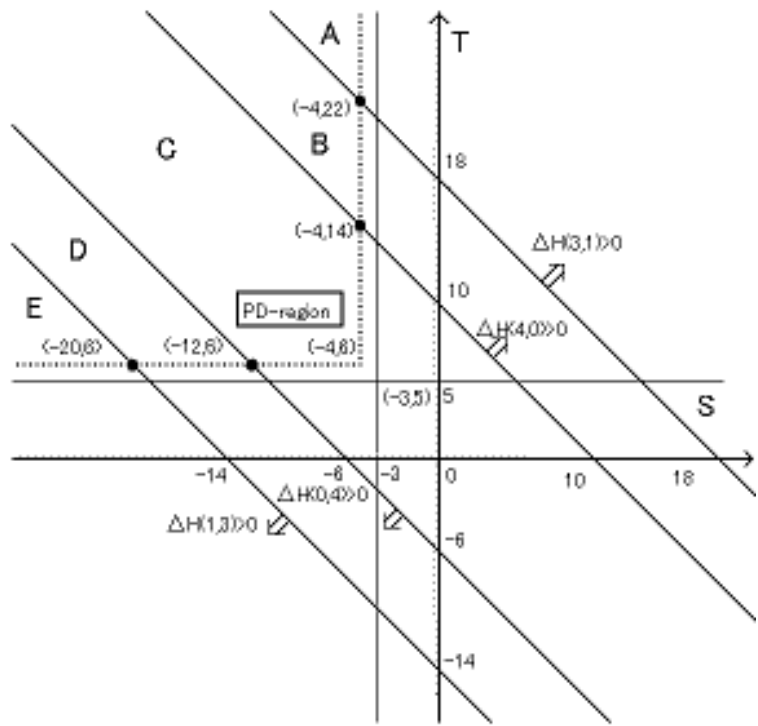}

\caption{Phase diagram}
\end{center}
\end{figure}

\section{Simulation results}
In this section the results of computer simulation on each reagion and the some disussions on them are obtained. We basically choose parameters according to Okada's book\cite{Okad}, i.e. $R=5,\; P=-3,\; T= 6,\; S=-4$ and the lattice size $n=13$ except for some special cases. 
The parameter $S$ and $T$, however, are changed to be some correct values such as identifying the corresponding region.
In the evolutionary PD, various types of initial configuration states are explored and many times experiments for a random initial state are made. 
The following initial configuration states are explored. \\
(a)Initial configuration has only defectors.\\
(b)Initial configuration has only defectors except one cooperator.\\
(c)Initial configuration has only cooperators except one defector.\\
(d)Initial configuration has only cooperators.\\
(e)Initial configuration has  cooperators and defectors at random positions where the ratio $\frac{N_1}{N_{-1}}$ of populations between cooperators and defectors is parametrized by $p (0\leq p \leq 1$).

Though the lattice size  $n=13$ seems to be too small, we have ascertained that essential results are unchanged by magnifying the lattice size except for some cases pointed specially.
   While states immediately converge in almost every case,  some interesting cases show rather complex behaviors such as chaotic-like behaviors.

\subsection{Selfish Evolution}
As all the agents follow this rule, full agents become defectors in the second step (i.e. after one round) because the dominant strategy in PD is a defection. Simulation results of course supports this argument. 

This situation is unchanged even in introducing a stohastic process by the Metropolis method with Boltzmann distribution, where player $i$ evade the above evolutionary rule with the probability $1-e^{-q\delta P_i}$, at low tempertature. 
($q$ is inverse of temperature.) 
  At high tempertature, however, a coexistence phase between cooperators and defectors appears. 
The critical value is $q=1.0 \sim 2.0$, that is, when $q>2$  the system is covered to defecrors after a few steps. 
Between $q=1.0 $ and $q= 2.0$ the system is unstable, and all defectors age and the age with a few cooperators revolve.
As $q$ decrease from 1.0 to 0, the average of the population of cooperators slowly increases to the half of total population with some fructuation. 
A typical example with $q=0.2$ is given in the Fig. 2.    
  
\begin{figure}[b]
\begin{center}
\includegraphics[width=\linewidth]{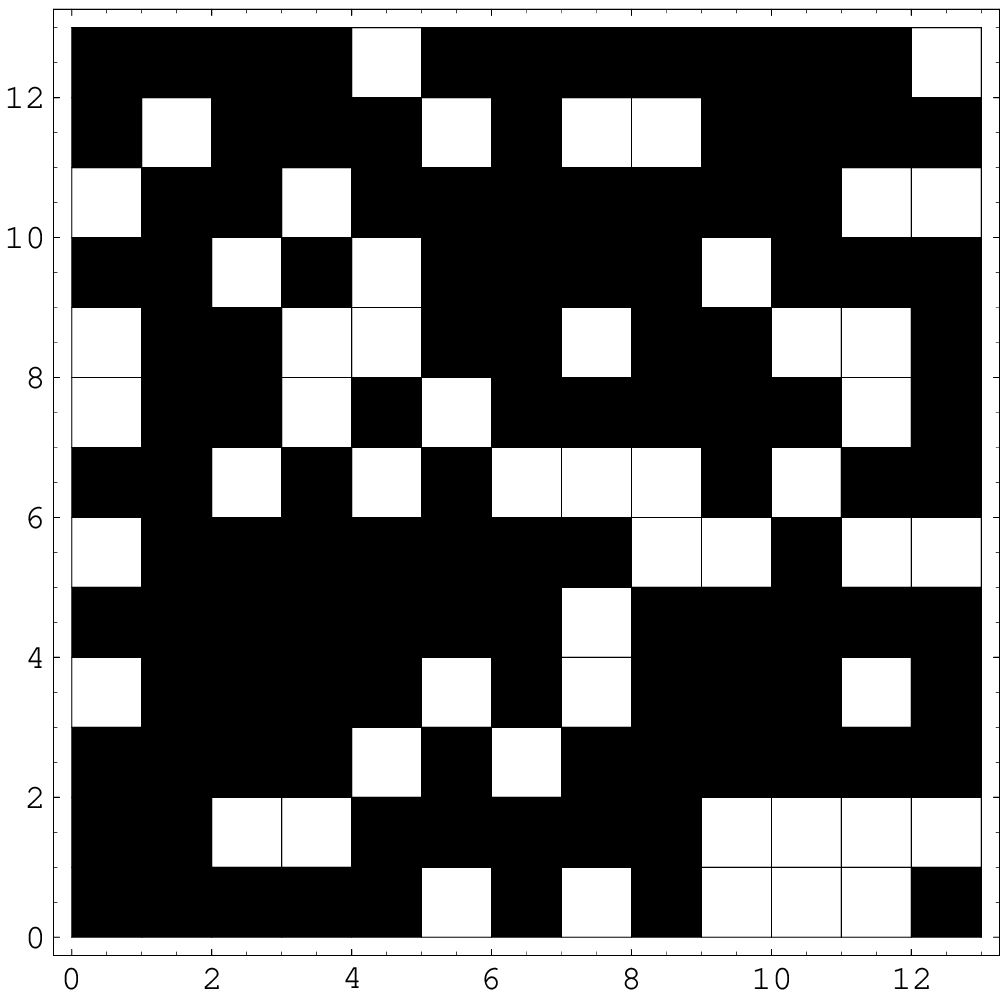}
\caption{Coexistence state diagram of C and D with q=0.2 where black lattices denote D-agents and white ones do C-agents.}
\end{center}
\end{figure}
These results also do not depend on the initial state of the system, starting even from  full defection (cooperation) initially.

\subsection{Globally Coupled Evolution}

(i)Region A\\
$\Delta H[4,0]>0,\; \Delta H[3,1]>0$ in this region. Thus an agent surrounded by many (3 and 4) C-agents becomes D-agent  but one surrounded by small number of C-agents remains C-agent at the next round.  The coexistence of C and D as the final state is predicted. 
 We simulate the case with $T=25$ and $T=50$. 
It turns out that independently of an initial state, the population ratio between C and D is $N_1:N_{-1}= 1:1$ within $3 \%$ error (for random initial state, within $10\% $ error) after a few rounds. A typical example is given in the Fig. 3.   
\begin{figure}[b]
\begin{center}
\includegraphics[width=\linewidth]{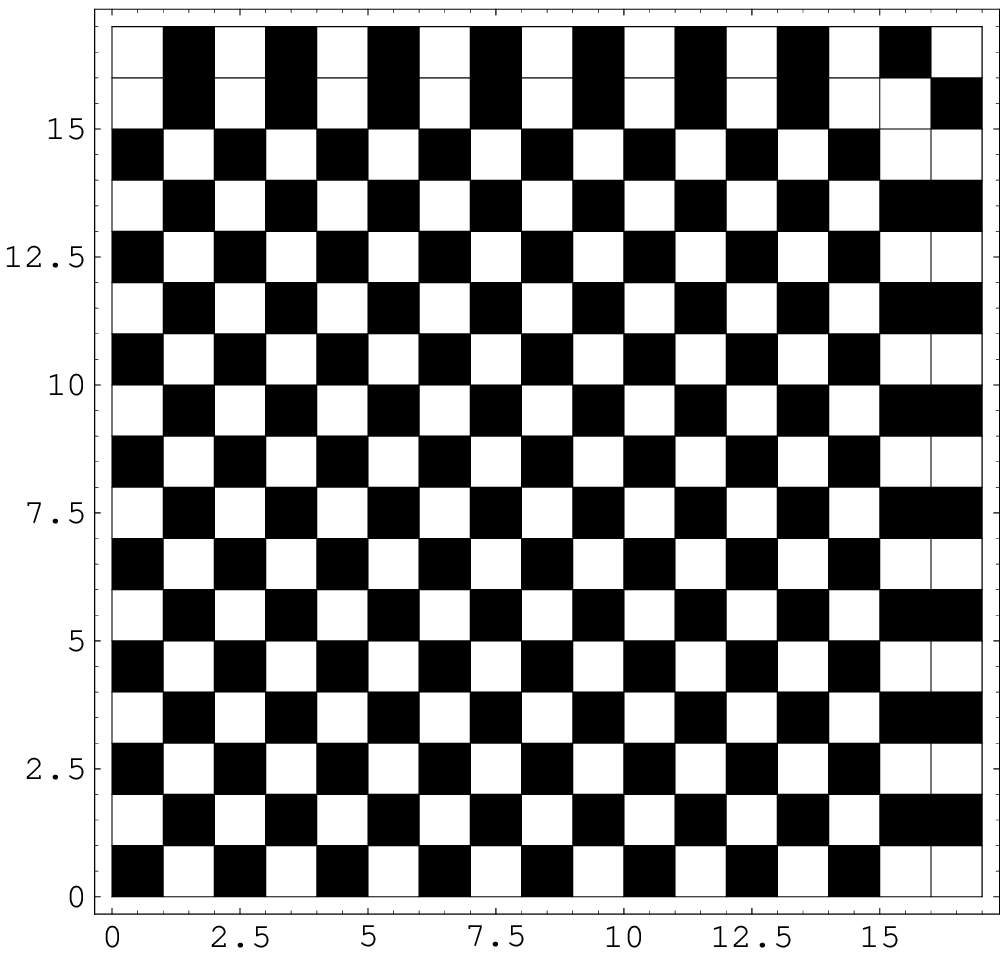}
\caption{Coexistence state diagram in the region A with initial (a) and $T=25$.}
\end{center}
\end{figure}

(ii)On the boundary between A and B\\
$\Delta H[4,0]>0,\; \Delta H[3,1]=0$ in this region with $T=22$. 
Agents take the same action as the region A, since we assume that an agent also change its strategy when $\Delta H=0$. 
Taking (a) and (b) initially, the state results in $N_1:N_{-1}= (a+n):a=(n+1):(n-1)$ with $n+2a=n^2$ by a simple analytic calculation from the resultant diagram after about $n$ rounds (Fig.4). Thus as $n \rightarrow \infty$, $N_1$ asymptotically comes close to $N_{-1}$
\begin{figure}[h]
\begin{center}
\includegraphics[width=\linewidth]{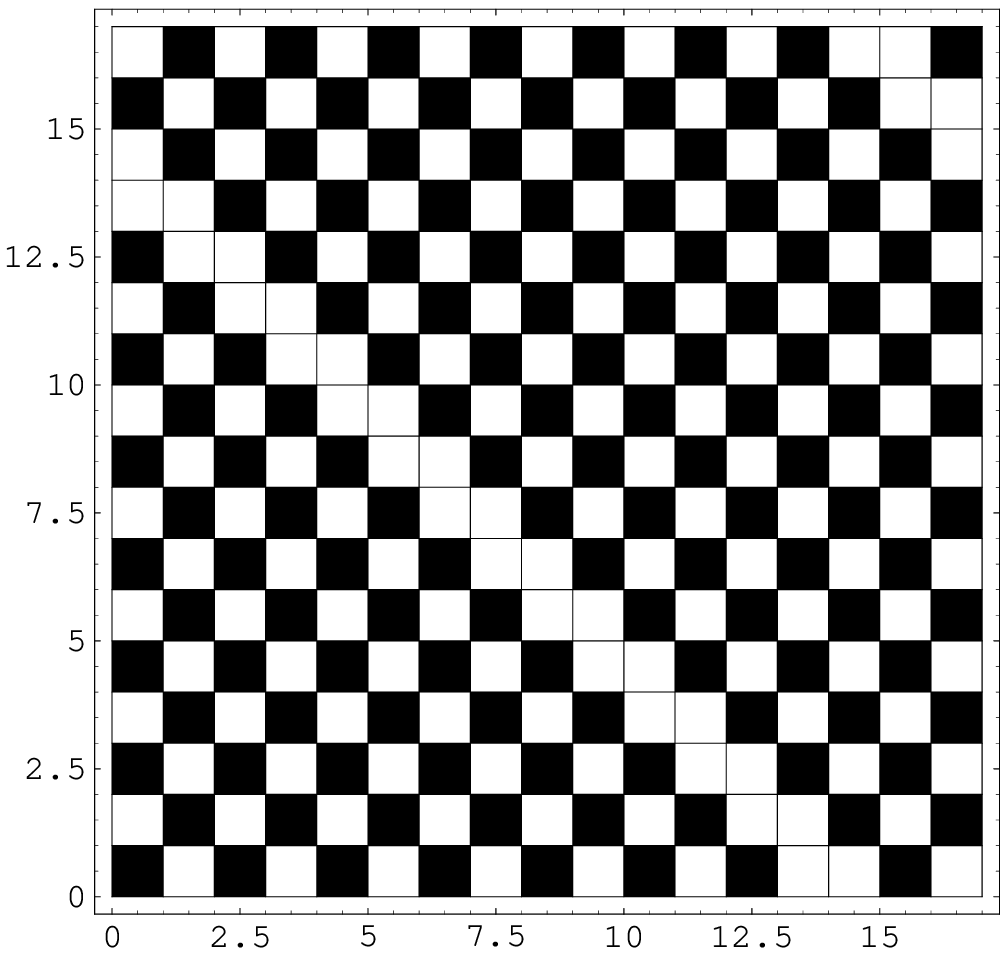}
\caption{Coexistence state diagram on the boundary between B and C with initial (c) and $T=22$}
\end{center}
\end{figure}

However this convergence is so unstable that the state does not converge even when initial state has two C-agents. 
Interestingly $N_1$ is exactly periodic and during the periodic age the total Hamiltoniani invariant, while the states itself is changing. 

 Random initial states with $p=0.5$ also expose almost convergent behaviour in $N_1$ but they are exactly periodic  really (Fig.5). The period is scale dependent. 
\begin{figure}[h]
\begin{center}
\includegraphics[width=\linewidth]{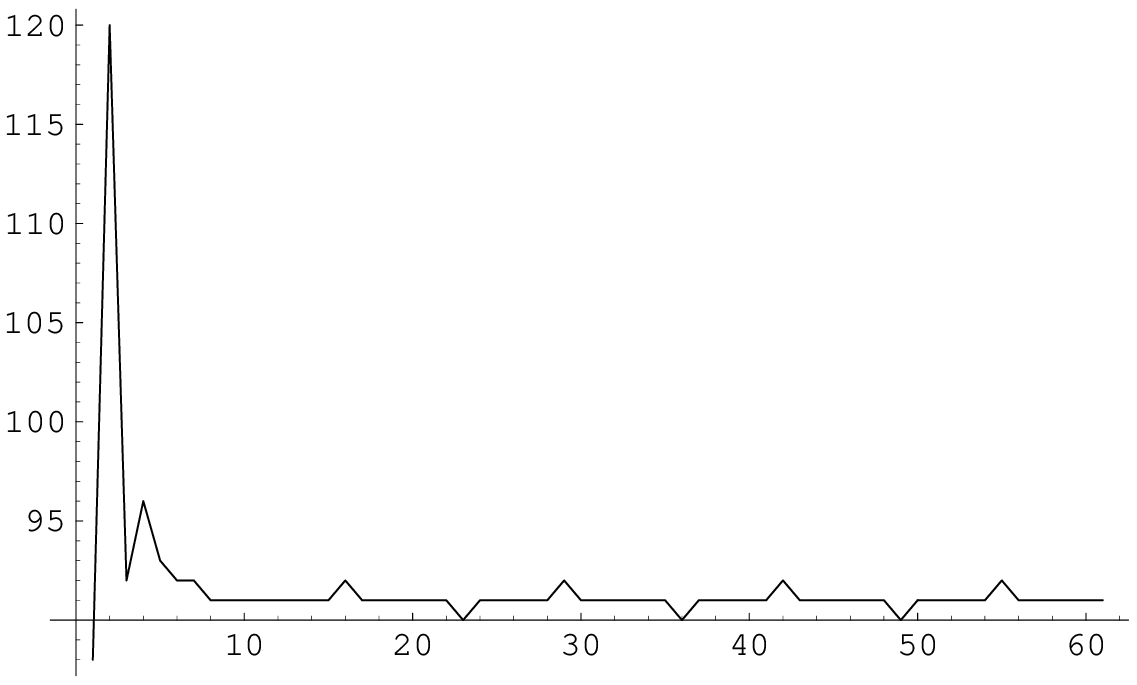}
\caption{The nunber of C-agents on the boundary between B and C with q=0.5 randam initial state and $T=22$}
\end{center}
\end{figure}

Generally various types of periodic behaviours in $N_1$ appear in random initial configurations but the configurations of the state  never converge and so are not periodic (of course strictly speaking, it is periodic because the dimension of the configuration space in size $n$ is $2^N$ ).   

For (d) and (c),  $N_C$ shows an intermittent but exactly periodic behaviour. The period is scale dependent and $n+c$ with the uncertain origin of the  constant $c\sim 14$ (fig.6). 
\begin{figure}[h]
\begin{center}
\includegraphics[width=\linewidth]{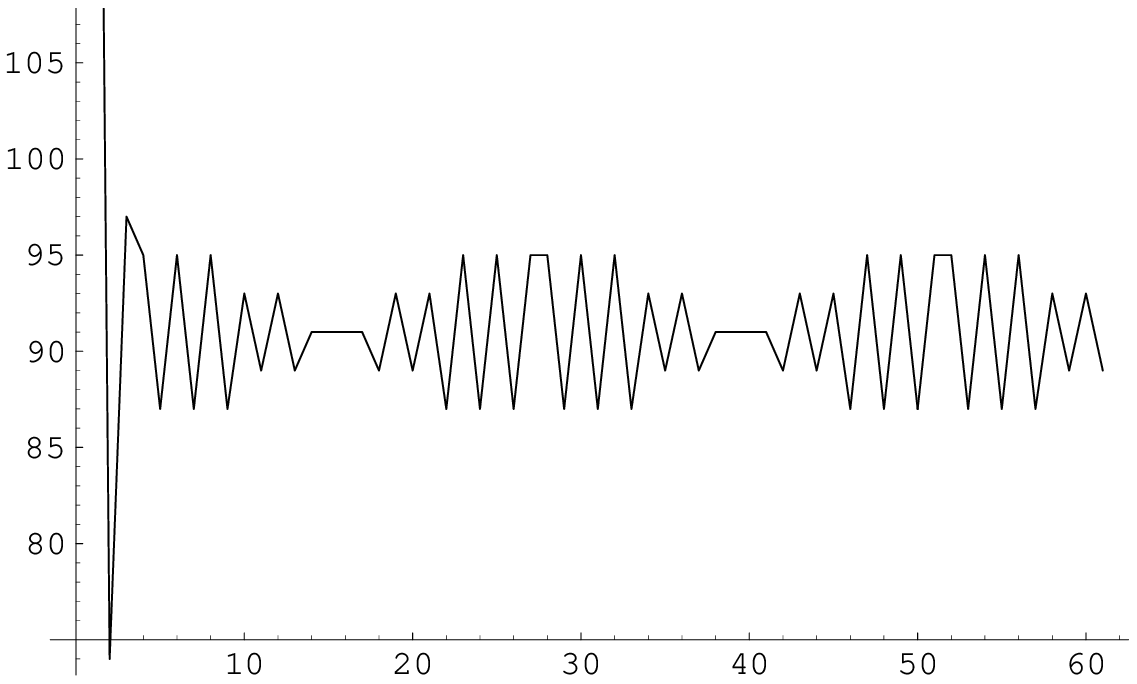}
\caption{The nunber of C-agents on the boundary between B and C with initial (a) and $T=22$.}
\end{center}
\end{figure}

(iii)Region B\\
In this region $\Delta H[4,0]>0$ and so  an agent surrounded by 4 C-agents only converts to D-agent. We choose $T=16$ as a typical value. For $n=13$ the state rapidly converge to be $N_C/N_D= 1.63$. By analytical consideration we easily find that 
$\frac{N_1}{N_{-1}} =\frac{(n^2 + 2n -3)}{(n^2 -2n+ 3)}\equiv r$ with $N_1 - N_{-1} = 2n-3$ and thus its ratio approaches to 1 as $n \rightarrow \infty$.  
 This result is common in the cases, (a),(b),(c),(d) as in the Fig.7, while the ratio fluctuates in the cases with randam initial configurations. Although the state at $ 0 \leq p< 0.2 $ and $0.8<p\leq 1.0$ converges to be the ratios 
$\sim r$, in the cases with $0.2 \leq p \leq 0.8$ the ratios are enhanced to the ratios  with $N_1 - N_{-1} \sim  4n$. 

\begin{figure}[h]
\begin{center}
\includegraphics[width=\linewidth]{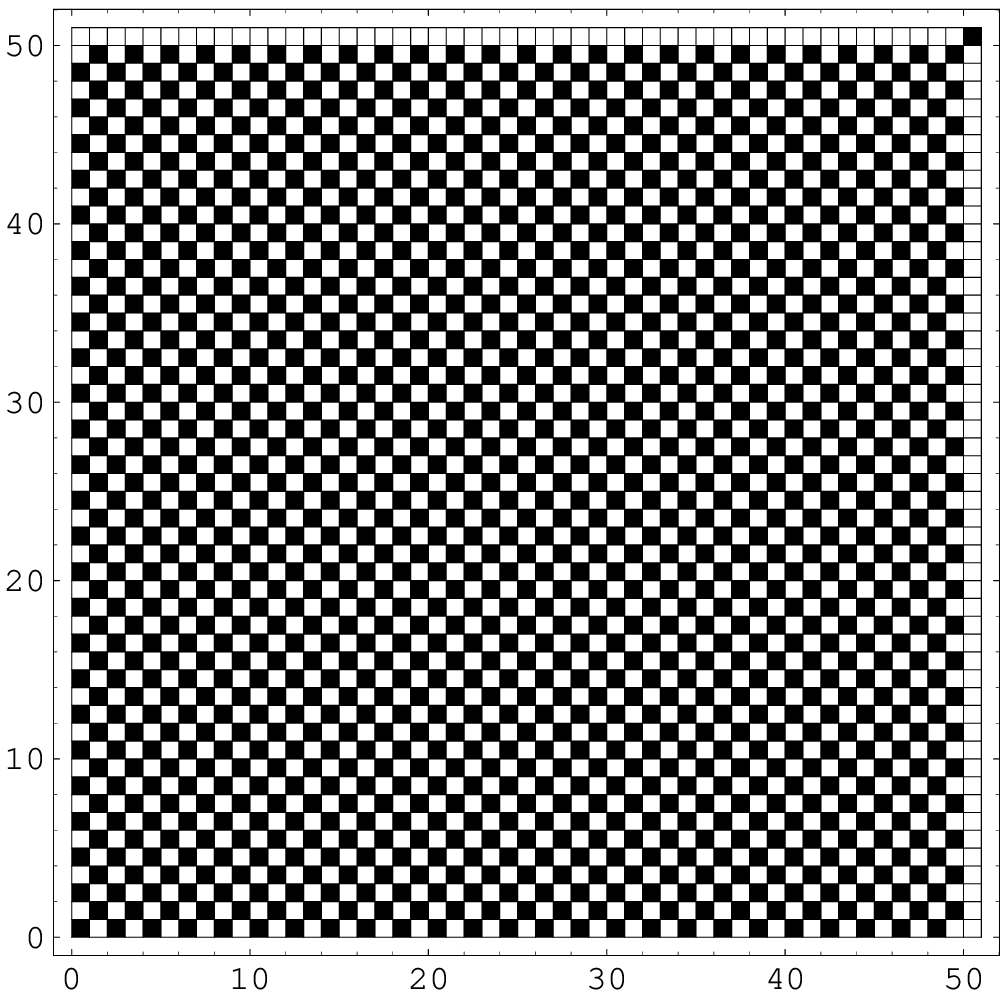}
\caption{Coexistence state diagram in the region B with $T=16$ and $n=51$}
\end{center}
\end{figure}

(iv)On the boundary between B and C\\
This region is identified by $\Delta H[4,0]=0$ with $T=14$ and all other $\Delta H<0$. Except for the cases with the randam initial confirurations, $N_1$ vibrates with round-dependent amplitude. 
Thus the amplitude itself vibrates with some long period (Fig.8), while the configurations show hightly  geometrical patterns.    
\begin{figure}[h]
\begin{center}
\includegraphics[width=\linewidth]{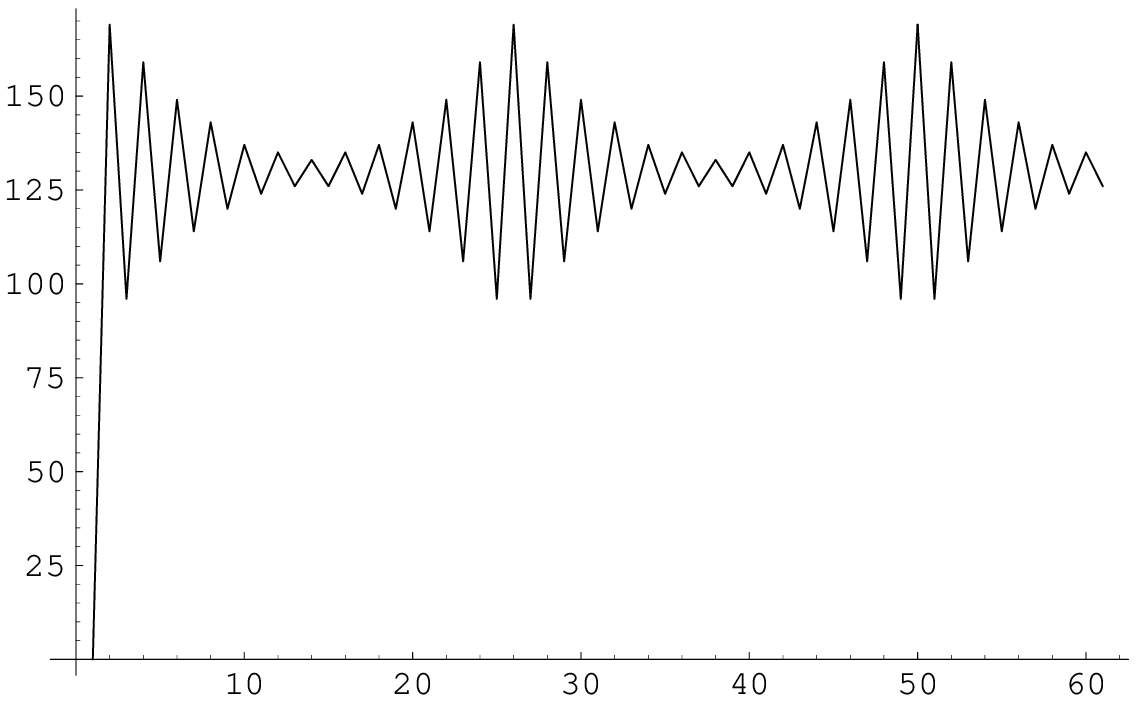}
\caption{$N_C$ sequence on the boundary between B and C with initial (a) and $T=14$}
\end{center}
\end{figure}

For $n=13$ the period is $23\pm 1$ (essentially $2(n-1)$) when states initially begin with (a),(b),(c),(d) and the total Hamiltonian remains invariant during   the period. On the other hand, begining with randam initial state, the evolution of the state is primafacie chaotic (Fig.9). However, constructing the returm map, we can find a weak correlation in $N_1$ data (the Fig.10). It is conjectured that the initial randamnes only weakly breaks the above periodic property.

\begin{figure}[h]
\begin{center}
\includegraphics[width=\linewidth]{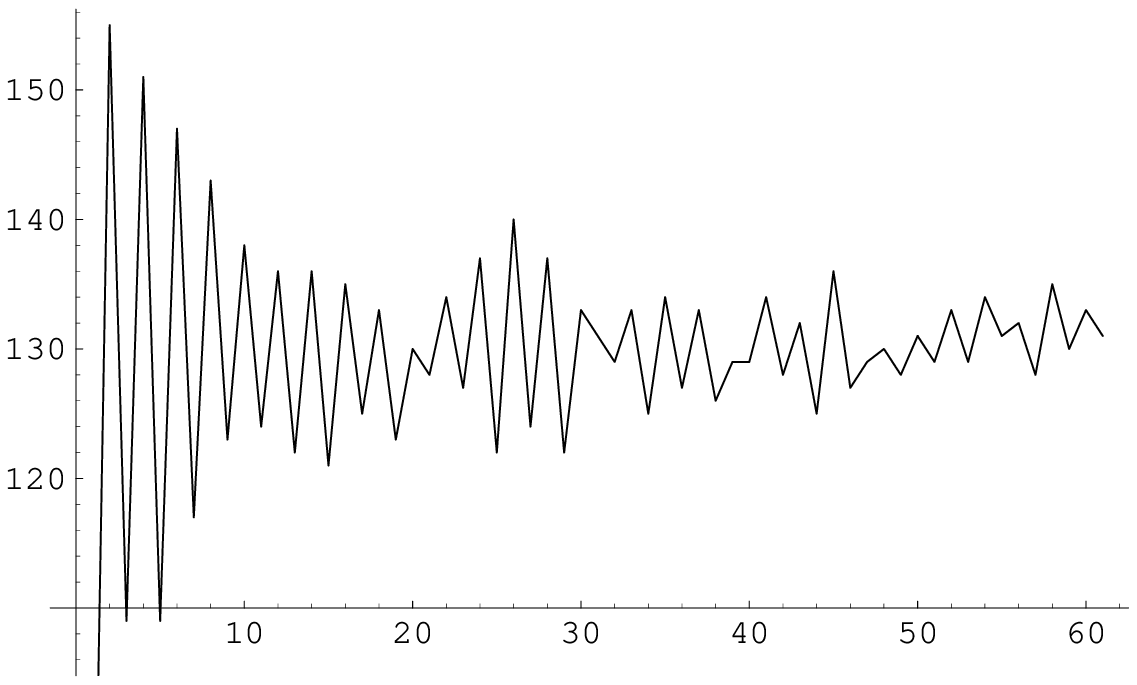}
\caption{$N_C$ sequence on the boundary between B and C in the random initial state with $p=0.5$ and $T=14$}
\end{center}
\end{figure}

\begin{figure}[h]
\begin{center}
\includegraphics[width=\linewidth]{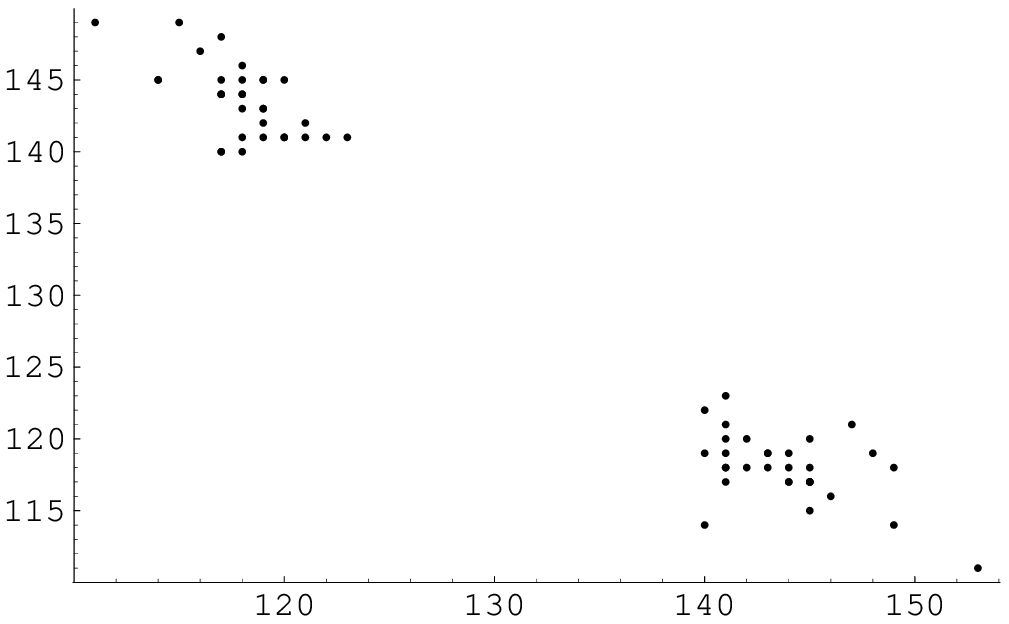}
\caption{Return Map on the region between B and C in the randam initial state with $p=0.5$ and $T=14$}
\end{center}
\end{figure}

(v)Region C (standard region)\\
In this case we have all  $\Delta H<0$. As a typical value of $S$ and $T$, $S=-4$ and $T=6$ are taken. As expected, the state rapidly converges to the state with all C, independent of initial configuration states. \\

(vi)On the boundary between D and C\\
Except for $\Delta H[0,4]=0$, all $\Delta H$ is negative. In this case the state rapidly converges to the state with all C with little marvel. 
This depends on the ordering of target agent as above mentioned. 
If C agent surrounded by 4 D-agents is lackily a first target in D dominant worlds, this situation will change drastically (this can be also confirmed by computer simulation really). In this simulation it is not so, that is, an agent D next to C agent first is remarked to turn to  C agent and so the original C agent can not become D agent when it is remarked.  Thus C agents only can live in the final state. 
This arises except for the case (a) where the state with all $R$ is stable. \\

(vii) Region D\\
In this region with $S=-18$ we have only positive $\Delta H[0,4]>0$. 
In this time an agent around without any C agent only becomes D agent. 
With the same reason as (vi) essentially,  C agents only can live in the final state, while the state (a) with all D agents anomalously is stable.  \\

(viii)On the boundary between D and E\\
In this case we take $S=-22$, and $\Delta H[1,3]=0$ and  $\Delta H[0,4]>0$. 
The phenomenological results are exactly same as the cases of (vii).\\

(ix) Region E\\
In this region with $S=-25$ we have two positive $\Delta H[0,4]>0$ and $\Delta H[1,3]>0$ . 
The phenomenological results are exactly same as the case of (vii) except one problematic point. 
In the case of (b), the state drops into the state with all D agents.
Since the randam initial state with $p=0.5$ finally comes to the state with all D agents, when the number of initial C agents decreases, there should be some kind of phase transition. We study this by varying the initial ratio $f_1$, that is to say, the parameter $p$.  They are summarized in the figure 11. 
Around $p=0.2$, we find that a weak phase transition occurs.

\begin{figure}[h]
\begin{center}
\includegraphics[width=\linewidth]{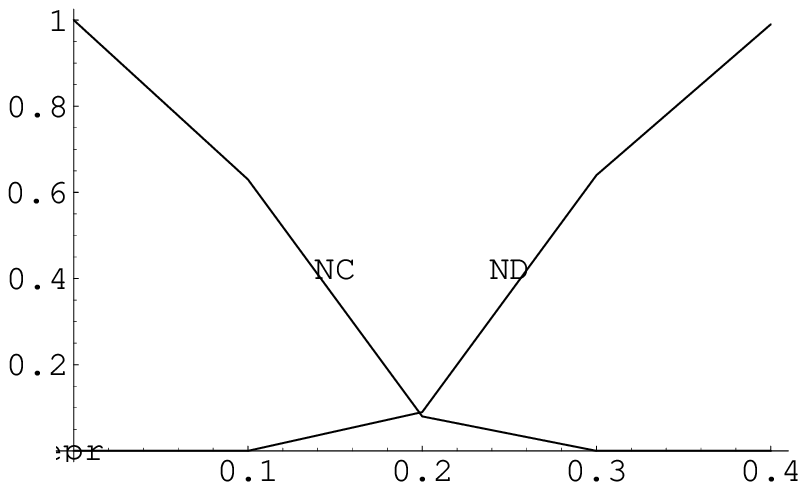}
\caption{$q$ vs. $N_C$ and $N_D$ diagram in the randam initial state with $S=-25$}
\end{center}
\end{figure}

\subsection{Some Remarks}
In the selfish evolution model without mutants, the state trivially comes to be full D-agent domitated world. In the case with mutants, however, there is a coextence phase between D and C. 
As varying the mutation ratio $q$, a sort of phase transition appears at around $q=1.5 \pm 0.5$. 
When $q<1.0$, it is inferred that the population ratio $N_1/N_{-1}$ is correlated with $q$, that is, the degree of randamness, from the simulation results, and thus the results may be almost consequential one.

In the globally coupled model, any interesting phenomena do not arise at critical points related to $\Delta H[1,3]$ and  $\Delta H[0,4]$, while rich structures appear at critical points related to $\Delta H[3,1]$ and  $\Delta H[4,0]$ which emerges at the regions breaking the condition (4). There are the coexistence phase in the latter regions, widely.   

This can be understood by considering the reduced payoff table which is constructed by adding the payoffs of two agents at each element of the payoff table 1.  This can be also derived by noticing that  $S$ and $T$ always appears pairwise and $P$ and $R$ always appears as multiples of 2 in $\Delta H$'s. 
The reduced table for PD is obtained in the Table 2.  
Basically agents tend to be a C-agent in the globally coupled rule (that is, (C,C) is favored) with the total payoff $2R$ from the table 2. 
This tendency is broken at $2R < T+S$ and the agents' preference is changed to 
 the  strategy D from C. 
This cosideration give the two effective critical points  related to $\Delta H[3,1]$ and  $\Delta H[4,0]$. 
On the other hand, $S+T<2P$ is irrelevant to the change of the tendency. 
Thus the critical points related to $\Delta H[1,3]$ and  $\Delta H[0,4]$ do not  contribute to this story drastically.

\begin{table}[h]\centering
\caption{Reduced payoff table for PD.  }
\begin{tabular}{|c|c|c|} \hline
  &\makebox[15mm]{C$_j$} & \makebox[15mm]{D$_{j}$} \\ \hline
C$_{i}$ & ($R+R$) & ($S+T$) \\ \hline
D$_{i}$ & ($T+S$) & ($P+P$) \\ \hline
\end{tabular}
\end{table}

To eliminate the order dependence of a target agent, we choose target agent at random to perform the same simulation. 
It, however, turns that the results are unaffected by the order essentially. The longer  period is merely needed untill the state converges.   
But, almost chaotic behabiour that does not show any regular behavior in the 2 and 3 dimensional embedding space appears in the region B with random initial state, which should be analyzed in details. 

\section{Summary}
In this article we propose two new evolutionary rules in which the explicit values in the payoff table play important roles apart from usual games. By these we mainly investigate the evolution of the spatio PD. 

In the selfish evolutionary role with mutations, coexistence phase appears with weak phase transition where the order parameter is $p$. This may be almost understood trivially. 

In the globally coupled rule it is natural to expect that a full D agent world is realized. As not expected, the rule presents rich structures in the spatio evolutionary PD. 
Analytically we find four critical points. When the condition $S+T<2R$ usually imposed is broken that is related two critical points, the coexistence phase appears in the wide range of parameter space, while the other two critical points do not any special role in the evolution (as expected all C world is generally generated), except for region E. 
The region E shows a little strange feature where there is a transition between all C world and all D world. 

Thus we uncover the entire phase structures under these rules. 
Detail analysis for some commplex data in $N_1$ such as chaotic behavior remain to us. 

\end{document}